# Review Article

# Nanocouplers for infrared and visible light

A. Andryieuski, and A.V. Lavrinenko

DTU Fotonik, Technical University of Denmark, Oersteds pl. 343, Kongens Lyngby, 2800 Denmark

Correspondence should be addressed to A. Andryieuski, andra@fotonik.dtu.dk

## Abstract

An efficient and compact coupler – a device that matches a micro-waveguide and a nano-waveguide – is an essential component for practical applications of nanophotonic systems. The number of coupling approaches has been rapidly increasing in the past ten years with the help of plasmonic structures and metamaterials. In this paper we overview recent as well as common solutions for nanocoupling. More specifically we consider the physical principles of operation of the devices based on a tapered waveguide section, a direct coupler, a lens and a scatterer and support them with a number of examples.

## 1. Introduction

Photonic components have advantages comparing to the electronic ones. Infrared and optical frequencies $10^{14} - 10^{15}$ Hz provide much broader operational bandwidth than the fastest electronic circuits. The losses in optical waveguides are smaller than in metallic wires. This is why, as D. Miller wrote, "the optical interconnects are progressively replacing wires" [1]. To achieve larger functionality on an integrated optical chip the optical components have to be miniaturized. A natural limitation, however, comes into play: the diffraction limit claims that we cannot focus light in a spot less than a half of the wavelength. The transverse size of conventional dielectric waveguides (for example, silicon waveguides) is also limited to a half of the wavelength. Only employment of metals allows to overcome the diffraction limit and to confine a wave to a smaller area, very often at the cost of increased propagation losses.

Nevertheless, the problem is not only to create efficient waveguides that provide subwavelength mode confinement, but also to make an efficient interface between free space or an optical fiber and a subwavelength nanowaveguide, that is to focus light and launch it efficiently into the waveguide. The artistic view of the situation is depicted in Fig. 1. Trying to pour water from a big bowl into a bottle with a narrow bottleneck, one would waste a lot. However, usage of a funnel simplifies the task and increases the efficiency significantly. An optical coupler plays the role of a funnel for light.

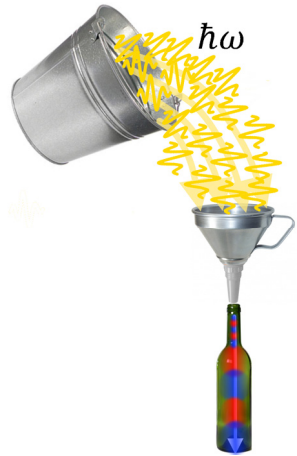

Figure 1. An artistic view of the problem of coupling light from a wide microscopic fiber to a nanoscopic waveguide. Employment of a coupler, which is represented by a funnel on the figure, minimizes the losses and simplifies optical alignment.

The problem of optical coupling originates from the pronounced modal mismatch between an optical fiber (a conventional single-mode telecommunication fiber has the core of 8 μm in diameter) and a nano-sized waveguide, which has a core less than 1 μm. They have a small overlap of modal fields that prevents from the efficient coupling. Long adiabatically tapered fibers can solve the coupling problem of a subwavelength waveguide, but for an efficient mode conversion the tapered region may reach the length of several millimeters that is totally incompatible with modern micro- and nano-fabrication. The compact efficient subwavelength couplers require new approaches.

We are witnessing a real explosion of various metamaterial and plasmonic solutions for focusing and nanocoupling in recent years. The reason for that are first of all the extraordinary optical properties the artificial metal-dielectric structures offer. Not only the proposed device geometries are different, but also the very physical principles they are based on vary significantly.

The goal of this paper is to give an overview of existing nanocoupling solutions for the visible and near-infrared (telecom) range and to reveal the most efficient ones. We wish to focus mostly on the physical effects employed for nanocoupling rather than on technical details of the devices. We are not going to provide the complete set of references on nanocouplers, since with heaps of articles already published and being published every month it is almost impossible. The cited papers should be considered rather as valuable examples of each physical effect employment.

The article is organized as following. In Section 2 the definitions of a nanocoupler and a nanowaveguide are given as well as the basic information on the common waveguide properties. Various physical principles, on which the nanocoupler realization can be grounded, are listed in Section 3. The tapered waveguide based nanocouplers are discussed in Section 4. Section 5 deals with the directional nanocouplers. The lenses based and scatterer based couplers are considered in Section 6 and Section 7 correspondingly. Other ideas are listed in Section 8. Conclusions summarize the paper.

## 2. Definitions

### 2.1. Nanowaveguide

We define a nanowaveguide as an electromagnetic waveguide that has lateral sizes in the "nano" range, i.e. 1 – 1000 nm. Depending on the waveguide material and the wavelength of light the nanowaveguide can be subwavelength or not. The modal size of the subwavelength waveguide is smaller than the size of a focused light beam can be. The nanowaveguides can be divided in two groups: dielectric and plasmonic waveguides. Their combinations, the so-called hybrid waveguides, are also possible.

### 2.2. Dielectric waveguide

The most common material for dielectric waveguides is silicon due to its high refractive index, transparency at the telecom wavelengths and CMOS compatibility. Typically, silicon waveguides are rib, ridge and photonic crystal waveguides. The high refractive index of the silicon waveguide makes it possible to reduce the cross-section down to 200x500 $nm^2$. The dielectric waveguide always has a cut-off size, below which the waveguide modes become leaky and the waveguide cannot transport light on reasonable distances.

Miniaturization of optoelectronic components requires decreasing the size of optical waveguides. However, the natural limitation comes into play. Due to diffraction the smallest size of an optical beam in a medium is on the order of the wavelength. If we consider a wave of the frequency $\omega$ propagating in the medium with the refractive index $n$, the wavenumber is $k = 2\pi n/\lambda$. The transverse wavevector components, for example, $k_x$ can take the values from $-k$ to $k$, so its maximal uncertainty can be $\Delta k_{x,max} = 2k = 4\pi n/\lambda$. Due to the uncertainty principle (or likewise Fourier transformations), the coordinate uncertainty $\Delta x$ is connected to the wavevector uncertainty $\Delta k_x$ [2] through

$$\Delta x \Delta k_x \geq 2\pi \qquad (1)$$

That limits the size of the light beam to

$$\Delta x = \frac{2\pi}{2k} = \frac{\lambda}{2n}. \qquad (2)$$

Considering, for example, the telecom wavelength $\lambda = 1.55$ μm and silica refractive index $n = 1.5$ we can estimate the smallest size of a light beam as $\Delta x_{min} = 517$ nm. Light can be confined to smaller spots only by using surface plasmons at an interface between metal and dielectric.

### 2.3. Plasmonic waveguide

Plasmonic waveguides are by default metal-dielectric waveguides. They attract a lot of attention since in some configurations they can show the absence of the cut-off wavelength at any waveguide size. Therefore, the mode size can be reduced to extremely small values but at the cost of increased optical losses. Another advantage of the plasmonic waveguides is the presence of metal that can be used not only as a waveguiding element but also as an electric contact that allows using it for tuning the dielectric surrounding (for example, due to the electro-optical or thermo-optical effects). Plasmonic waveguides are currently considered for the potential replacement of the electronic interconnects in the future generation integrated circuits.

Surface plasmon polaritons (SPPs) are the eigenmodes of a metal-dielectric interface. SPPs are combined light – electrons density waves. On a flat metal (permittivity $\varepsilon_1$) – dielectric (permittivity $\varepsilon_2$) interface the SPPs are transverse waves with the magnetic field parallel to the interface. In the visible and near-infrared ranges the permittivity of metal is very dispersive. The propagating SPP solutions correspond to a case Re($\varepsilon_1 + \varepsilon_2$) < 0. For a detailed background on plasmonics theory and applications we refer the reader to [3], [4].

Several types of plasmonic waveguiding structures were proposed, e.g. metal-insulator-metal and insulator-metal-insulator multilayered structures [5], strip [6], trenches and V-grooves [7], [8], wedge [9], slot [10], [11] and nanoparticles chain [12] waveguides. A comprehensive overview of the plasmonic waveguides and nanoplasmonic systems can be found in [3], [4], [13–15].

## 2.4. Nanocoupler definition

A nanocoupler is a device that facilitates coupling of light from free space or a macroscopic optical fiber to a nanowaveguide. It can be understood as a focusing device, a sort of an optical funnel (see Fig.1) that squeezes light into a small spot. If we talk about matching a thick and a thin waveguides that have different modal distributions, the nanocoupler can be also understood as a mode convertor, a device that transforms a mode of the thick waveguide to a mode of the thin waveguide (see Fig. 2).

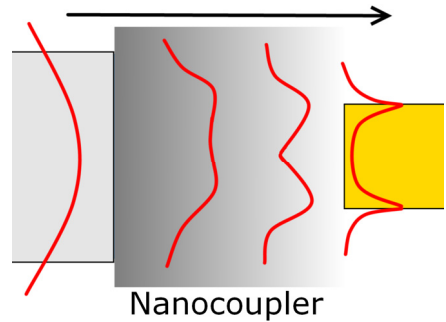

Figure 2. Nanocoupler concept: a focusing device or a mode convertor. As an example the mode of the dielectric waveguide (left) is gradually transformed into the mode of the plasmonic waveguide (right).

## 2.5. Requirements to a nanocoupler

We understand the nanocoupler as a focusing device or a mode convertor that gives a high coupling efficiency (CE), which is the ratio of the power loaded into the waveguide $P_{WG}$ to the power incident on the nanocoupler from free space or power delivered by a thick waveguide $P_{inc}$

$$CE = \frac{P_{WG}}{P_{inc}} \qquad (3)$$

High coupling efficiency automatically means low losses, including absorption $A$, reflection $R$ and scattering $S$. Additional requirements to the nanocoupler can be:

1. Small size on the order of several micrometers. This is an important requirement for the integration of the nanocoupler within an optical integrated circuit.

2. Simplicity and small price of fabrication. This requirement is important for the mass production but not so important on the stage of the scientific development.

3. Spectral selectivity. The importance of this property depends on the application. In many cases a broad bandwidth is desirable. However, there are some applications, where a narrow bandwidth is preferable, for example, if the coupler is used at the same time for the wavelength demultiplexing.

4. Polarization sensitivity. The importance of this requirement depends on the application and on the selected nanowaveguide. If the nanowaveguide is polarization sensitive, the nanocoupler's working polarization should be matched with the polarization of the waveguide or it should be polarization insensitive.

We have to emphasize that the nanocoupler differs from a nanofocusing or a nanoimaging device. The goal of nanoimaging or nanofocusing is only to image a tiny light source or focus light into a small spot, respectively, no matter how large is the fraction of the transmitted power with respect to the incident power, while for the nanocoupler the coupling efficiency is the most crucial parameter. For example, the stimulated emission depletion technique [16], [17] allows for a fantastic resolution below 10 nm in the visible light imaging systems, but it can hardly be used for an efficient coupling.

## 3. Nanocouplers classification

The role of the nanocoupler is to match the impedances (or wavevectors) and field profiles of an incident wave and a mode of the accepting nanowaveguide. Depending on the employed physical mechanisms several types of nanocouplers can be singled out:

1. Tapered waveguides (Fig. 3(a)) which will be referred to as *tapered waveguide coupler*. In this case a waveguide with the gradually reduced core cross-section compresses a wave towards the matching parameters with the accepting nanowaveguide.

2. Light can be coupled first to a wide waveguide (for example, a silicon or long-range surface plasmon-polariton waveguide) and then with the help of a directional coupler or a resonant stub coupled into a smaller nanowaveguide (Fig. 3(b)). We will refer this case to as a *direct coupler*.

3. Light can be tightly focused with a lens (Fig. 3(c)). This is what we are referring to as a *lens coupler*.

4. A single or multiple scatterers can be used for coupling (Fig. 3(d)). Correspondingly, the device is called a *scatterer coupler*.

5. There are some other ideas that do not fall into the abovementioned categories. We will refer to them under a unified shield as *other solutions*.

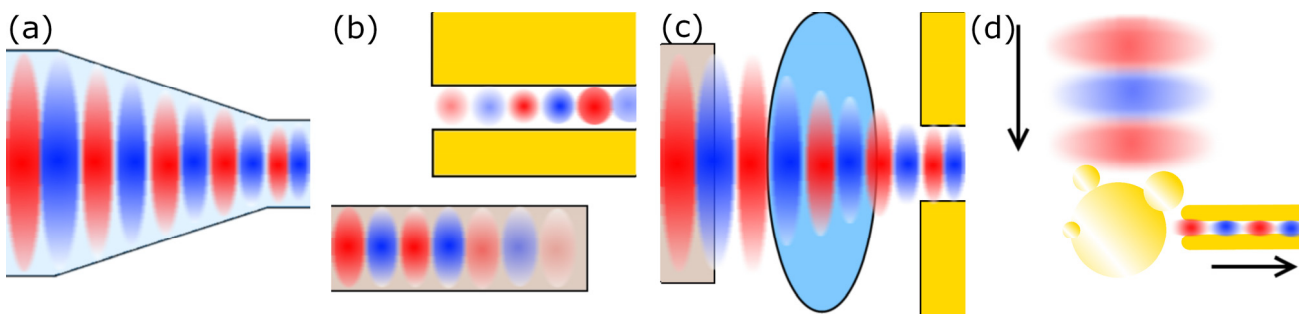

Figure 3. Types of the possible nanocouplers realizations: (a) tapered waveguide coupler, (b) direct coupler transferring the power from a wide waveguide to a narrow waveguide, (c) lens coupler, (d) scatterer coupler.

The nanocoupler can be arranged as a separate device (Fig. 4(a)), as a device integrated with a waveguide on a chip (Fig. 4(b)) or as a device integrated with an excitation fiber (Fig. 4(c)). For the practical applications the integrated configurations (b) and (c) are preferable, since a lesser number of movable parts simplifies optical alignment.

Two main excitation configurations are important in the practical applications: lateral coupling (Fig. 5(a)), when the incident wave direction coincides with the nanowaveguide; and vertical coupling (Fig. 5(b)), when the light direction is perpendicular to the nanowaveguide. The second configuration can provide not only in-coupling from free space, but also communication of the optical elements between two layers of an optical integrated circuit. Incident angles of light different from 0 and 90 degrees are also possible.

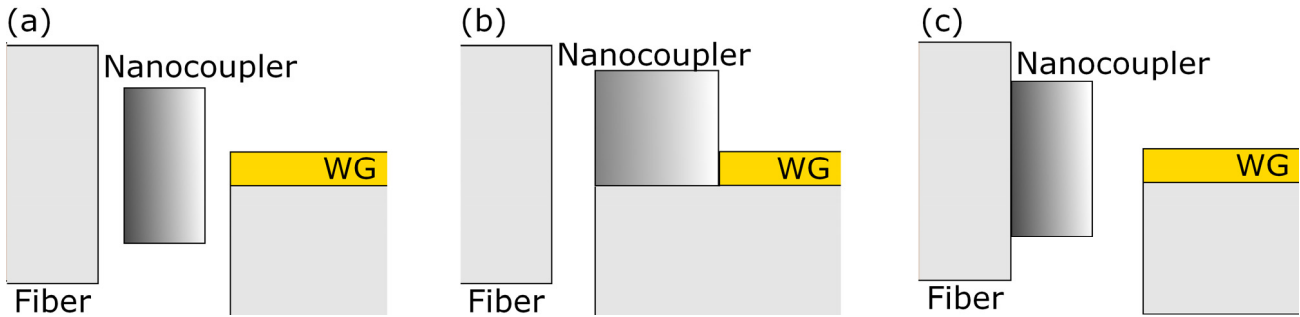

Figure 4. Possible geometrical configurations of the coupler: (a) a separate device, (b) integrated with a nanowaveguide (WG) on the same chip, (c) integrated with an excitation fiber.

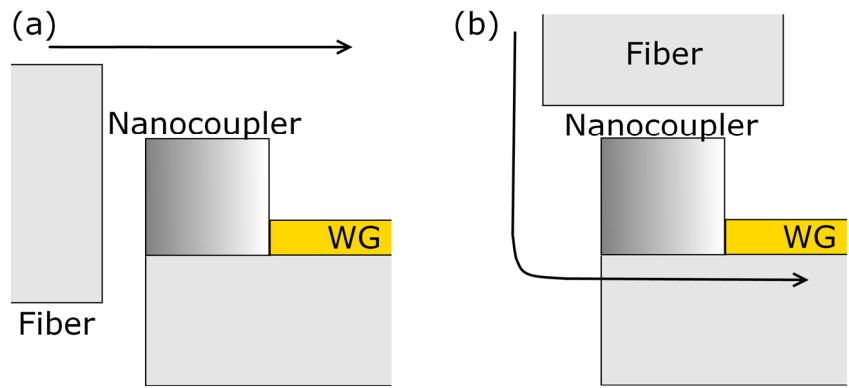

Figure 5. Two practically important coupling configurations: (a) lateral and (b) vertical. Black arrows show the direction of light propagation.

## 4. Tapered waveguide coupler

As we already mentioned the tapered waveguide coupler is nothing else but a waveguide with the gradually decreasing core. A mode of the waveguide is subjected to adiabatic compression, when propagating to the tip. Thus it reaches more favorable conditions for coupling to the nanowaveguide. The tapering angle and the rate of compression determine the coupling efficiency and depend on the properties of materials. Therefore, further classification of the tapered waveguides follows the material properties:

1. Fully dielectric tapers having the dielectric core and the dielectric cladding (Fig. 6(a)).

2. Hybrid tapers having the metallic core and dielectric cladding or the dielectric core and metallic cladding (Fig. 6(b)).

3. Metamaterial tapers having the structured metal-dielectric composite core (Fig. 6(c)).

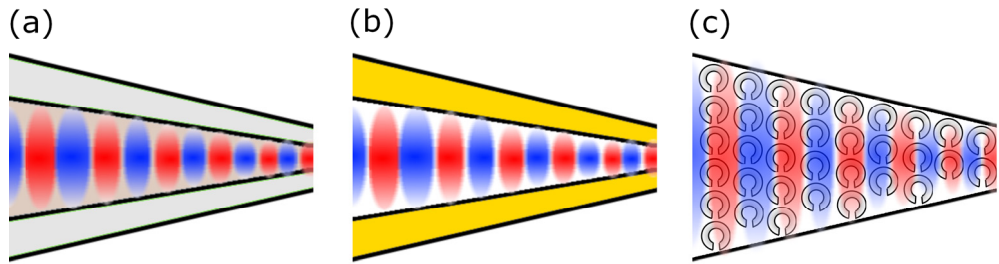

Figure 6. The types of taper couplers: (a) fully dielectric taper, (b) hybrid taper with metal core or cladding, (c) metamaterial taper.

### 4.1. Dielectric core and cladding

If one takes an optical fiber, heats it up and then pulls, the diameters of the core and shell simultaneously decrease. So it is possible to draw the core down to a nanosize diameter. Low-loss tapered fibers were fabricated and measured [18]. The optical losses for a fiber of diameter $d = 750$ nm at $\lambda = 1.55$ μm are 0.017 dB/mm. However, it was also theoretically shown [19] that even an ideal fiber does not allow the substantial core thinning, since the propagating mode completely vanishes for the core size of one order of magnitude smaller than the wavelength. In the real fibers with imperfections this limitation is even tougher.

Nevertheless, the dielectric tapered waveguide coupler finds useful applications. As L. Zimmermann reported at the Silicon Photonics Workshop in 2011 [20], an inverted taper is commonly used for lateral coupling from a high numerical aperture lensed optical fibers with a spot size of 3 μm to a 200x500 nm$^2$ silicon waveguide. The inverted taper is also used for coupling to and from the slow light photonic crystal waveguide [21] reducing the insertion loss value to 10 dB. Tapering came out to be useful to out-coupling efficiency increase and far-field shaping of the nanopillar single photon sources [22], [23]. The coupling efficiency can be very large (80% and higher) for a certain pairs waveguides, for example for silicon ridge and silicon slot waveguides [24], [25].

### 4.2. Metallic core or cladding

Tapered fiber tips covered with metal are the essential part of the scanning near-field optical microscopy. It allows spatial compression of light below the diffraction limit by employment of SPPs [26]. Metal-dielectric-metal tapered plate waveguides were theoretically and experimentally demonstrated for the terahertz [27–30] and optical ranges [31–34] showing the ability of ultrahigh energy concentration. In the latter work [34] the maximal transmittivity of 8% was measured for the wavelength $\lambda = 780$ nm in an adiabatically tapered metalized fiber with the tip opening of about 150 nm. The prominent property of such system is that a metal-dielectric-metal waveguide does not have a cut-off, and only losses limit its performance. Nanofocusing with a V-shaped metallic groove, which is a similar system, was experimentally demonstrated [35]. A wave with $\lambda = 1.5$ μm wavelength was focused into a spot of $\lambda/40$. The reported energy efficiency was 50%. In 2004 M. Stockman predicted that a tapered metallic wire allows the giant energy accumulation and intensity enhancement at the tip of the metal taper [36]. The idea is applicable not only in the optical range but also in the terahertz range if the surface of the tapered wire is corrugated [37]. An experimental realization of the tapered wire structure was reported [38]. In this case light with $\lambda = 1.55$ μm was coupled to a 2000 nm wide wire and compressed down to 90 nm. The transmittivity of 20% was shown.

Another approach was proposed in [35]. A metallic funnel consisting of long metallic nanocylinders of the radii from 10 to 24 nm separated with a 2 nm thick dielectric was shown to focus a Gaussian beam of 200 nm full width into a spot of 20 nm with a power transmittivity of 80% [39]. However, a practical realization of such system is hardly possible at the moment.

### 4.3. Metamaterial core

The idea behind this nanocoupler type is to use a metamaterial as the taper core. The advantage of utilization of metamaterials is that their properties are not strictly limited to that we have from natural materials and can be designed to reach amazing diversity and values. An optical funnel containing a metal-dielectric photonic crystal was proposed in [40]. The field compression down to $\lambda/30$ with a transmission of 13% was reported. The metamaterial based nanotips for field enhancement were theoretically proposed [41], [42]. The nanotip consists of metallic nanospheres of a gradually changing density. They can be used for field concentration (200 times field enhancement was reported) and the light compression to a spot of 10 nm. In the work [41] it is stated that the power efficiency of such metamaterial tip is larger than of a metalized taper with the 10 nm output hole, but the exact values of the transmittivity are not mentioned.

### Summary

A dielectric adiabatically tapered fiber is a well-known instrument of field concentration. However, it can provide either small spot size with low transmission or high transmission at the cost of large size of the light beam and the coupler itself. The metal or metamaterial based tapers can be of significantly smaller length. They can give field concentration and reasonable transmittivity (e.g. 20%). The difficulties in practical realization of effective design solutions are connected with the technological constrains, for example, fabrication of a regularly packed bundle of metallic cylinders of radius 10 nm separated with 2 nm of dielectric is currently out of reach.

## 5. Direct coupler

Based on the geometrical position of one waveguide with respect to another, the direct couplers can be divided into the following classes:

1. One waveguide next to or inside another (Fig. 7(a)).

2. End-fire direct connection of the waveguides (Fig. 7(b)).

3. Resonant stub between the waveguides (Fig. 7(c)).

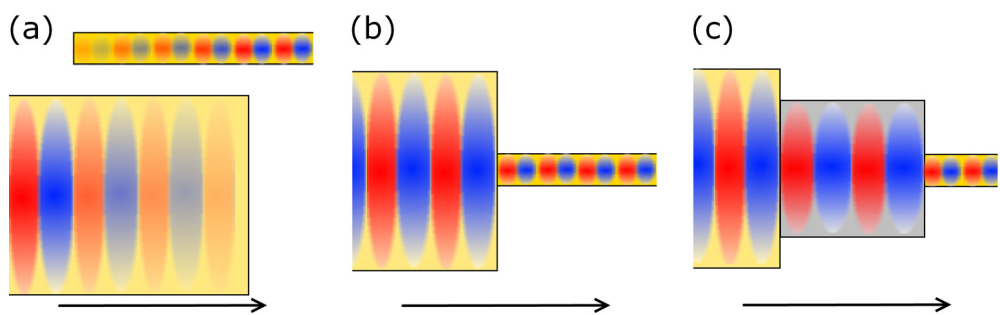

Figure 7. Types of the direct couplers: (a) one waveguide next to or inside another, (b) end-fire coupling, (c) resonant stub between waveguides.

Direct couplers as we classified them are either directional couplers or schemes including end-fire coupling. Directional couplers are classical components of photonic integrated circuits. They are typically used for wavelength division or light switching. The principle of their operation is based on the coupling between parallel waveguides due to overlapping of their modes fields. The coupling leads to the hybrid modes (supermodes, compound modes) formation. The strength of the mode coupling can be controlled by the distance between the cores. Therefore, such couplers require preliminary feeding of a large waveguide at an intermediate step. Then disposing the first waveguide in a close proximity with a nanowaveguide we initiate their coupling with a consequent transit of the light energy in the second channel. Directional couplers are characterized by the coupling length, – the distance, on which the maximal amount of the energy is transferred in the second channel and vise verse. The coupling length, in turn, is proportional to the wavenumbers mismatch. As a rule, directional couplers have a relatively large coupling length, what results in the total size of such systems about 10-20 µm.

The end-fire coupling scheme also exploits the fields overlapping mechanism. Here the coupling appears as a result of the field profiles matching. The scheme does not require a lengthy coupling part, however, short sizes come at a price of lower coupling efficiency. To improve the efficiency a resonant stub is often placed between the input and nanosized waveguides.

### 5.1. One waveguide next to or inside another

A typical scheme for direct coupling is to arrange one waveguide above another such that the eigenmodes of the waveguides hybridize, and the waveguides become coupled. In such case energy from the input mode is transferred from one waveguide to another and back. Making the overlap region equal to the coupling length it is possible to achieve the maximal energy transfer efficiency. Such systems were proposed, simulated and experimentally realized for dielectric [43] and plasmonic long- and short-range waveguides [44–48]. The coupling length may vary from several micrometers [47], [49] to hundreds of micrometers [43]. For certain waveguides (dielectric to long-range surface plasmon polariton) the coupling efficiency can be extremely large. The theoretical prediction of coupling efficiency CE= 60% from dielectric to plasmonic slot waveguide at λ=1.55 µm was confirmed in measurements [47]. However, we should take into account that light should be first coupled from a fiber to an input (silicon) waveguide and that the maximal coupling efficiency is in the order of 60-70%. That makes the total coupling efficiency of the system about 30-40%. It is possible to obtain light coupling by placing metal nanowires on top of the dielectric waveguide perpendicular to the latter [50]. In such arrangement the overlapping region is extremely small, and the coupling efficiency of such system is reported to be only 1%.

The configuration when one waveguide protrudes into another is possible to certain waveguide types only, since the bigger waveguide should contain empty space inside. For example, it is possible to insert a silicon waveguide inside a plasmonic slot waveguide, but not into another silicon waveguide. Several designs for telecom wavelengths were proposed [51–53] and experimentally realized, giving a theoretical coupling efficiency of 88% and measured coupling efficiency about 35% [53].

### 5.2. End-fire coupling

The end-fire coupling is the simplest case of the waveguides connections and it is common in the optical communication systems. Nevertheless, the end-fire coupling from a dielectric to plasmonic slot waveguide or a nanowire is not usually very efficient due to the pronounced impedances mismatch. However, if the optimization of the geometry is conducted the coupling efficiency more than 70% [54], [55] and even 90% can be achieved [56], [57]. For example, at the wavelength of λ = 1.55 μm the coupling efficiency of 80% was measured for a long-range surface plasmon polaritons waveguide connected to a silicon waveguide [58]. In another work the coupling from a silicon to plasmonic slot waveguide with 30% efficiency at λ = 1.55 μm was experimentally demonstrated [59]. End-fire coupler based on optical tunneling can be used for identical waveguides separated with a small gap [60]. Varying the width of the gap one can tune the transmission and reflection in a wide range.

### 5.3. Resonant stub between the waveguides

The use of resonant stubs (for example, a λ/4-transformer) for the efficient wave coupling is the well-known technique in the microwave waveguides engineering. It is based on the resonant transmission increase due to the constructive interference similarly to the antireflection coatings for lenses. The same concept can be used in the optical range, for the plasmonic waveguides of different cross-sections [11], [61–63] and for the silicon and plasmonic waveguides [64] matching. The coupling efficiency can be enhanced in comparison with the end-fire coupling. However, as based on the resonant phenomena the scheme has limited bandwidth and very individual application range.

The λ/4-transfomer matching 500 nm and 50 nm wide plasmonic transmission lines with the coupling efficiency of 86% was shown numerically [61]. Matching of a 300 nm wide silicon waveguide with a 40 nm wide plasmonic slot waveguide with the coupling efficiency of 88% was also demonstrated [64].

### Summary

While giving high values of the coupling efficiency (in case of a directional coupler up to 60%) and being feasible for fabrication, the direct nanocouplers require additional structures such as an input silicon waveguide and a preliminary coupler to this silicon waveguide. That makes the total CE of up to 40% and extends the nanocoupler dimensions up to several dozens of micrometers. Moreover, additional structures (for example, a preliminary coupler) may require additional processing steps during fabrication.

## 6. Lens coupler

Lens is a well-known focusing device. However, we should emphasize that the requirements for a nanocoupler are stricter than to a focusing device, since nanocoupler should provide, apart from focusing, the high coupling efficiency. To be implemented as a nanocoupler a lens must have high transmission and execute matching of the focused beam to the nanowaveguide mode.

Based on the lens material and their functionality we divide lenses into the following categories:

1. Dielectric lens (Fig. 8(a)).

2. Plasmonic lens (Fig. 8(b)).

3. Negative refractive index lens (Fig. 8(c)).

4. Photonic crystal lens (Fig. 8(d)).

5. Hyperlens (Fig. 8(e)).

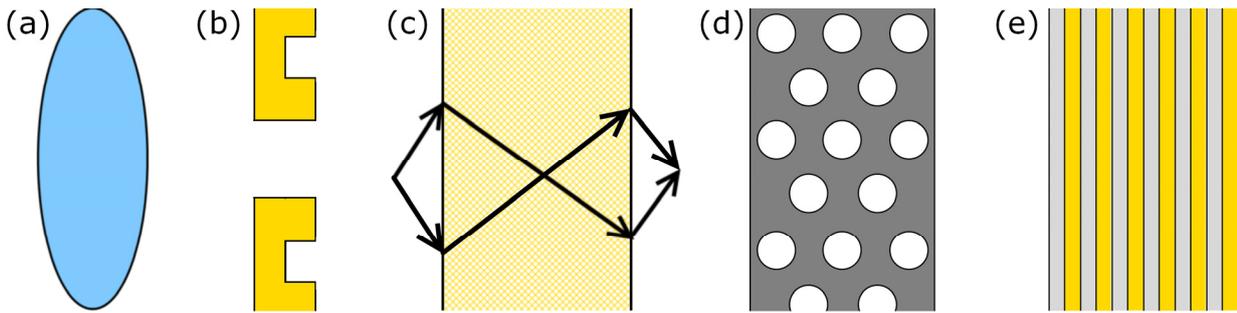

Figure 8. Types of lens couplers: (a) dielectric lens, (b) plasmonic lens, (c) negative index lens, (d) photonic crystal lens, (e) hyperlens.

## 6.1. Dielectric lens

A dielectric lens is a focusing device known for several centuries. It is well described in classical optical textbooks (see, for example [65]). The lens can provide the excellent light transmission, but its resolution is diffraction limited and for coherent light cannot be better than $0.77\lambda/NA$, where $NA$ is a numerical aperture. The numerical aperture cannot exceed the refractive index $n$ of a surrounding material. Even such resolution is hard to achieve in practice, as it requires a complex optical setup with a high numerical aperture objective; therefore, a conventional dielectric lens is not a suitable solution for nanocoupling.

Except a standard dielectric lens we should mention also two other focusing devices, namely, a Fresnel lens and a graded index (GRIN) lens. The Fresnel lens and zone plate are well described in optical textbooks [65]. The zone plate consists of a set of concentric rings (Fresnel zones). Rings are transparent and non-transparent in the alternating order or have the 180° phase difference for the transmitted light. An important requirement is that the geometrical sizes of the zones impose the constructive interference conditions in the desired focal point. Being much lighter and much more compact than a bulk dielectric lens, the Fresnel lenses have worse resolution due to the diffraction at the zones' borders.

The GRIN lens was developed for photonics packaging [66]. It is called a lens even though its shape is far from being concave or convex (etymologically "lens" means to be of a double convex shape). It is usually a multilayer dielectric structure with a gradually changing refractive index. The refractive index gradient forces the wave propagating along the waveguide to refract and concentrate in a thin bottom layer, from where it is easier to launch it into a smaller waveguide. For example, a GRIN lens for silicon waveguides [67] allows for        coupling efficiency.

A very useful for practical purposes is a focusing optical fiber that combines mechanical flexibility of a fiber with the focusing effect of a lens. Commercial focusing fibers can provide a focal spot down to 2 μm in diameter [68]. The focusing effect of the fiber can be reached either by making a lensed output end or by the gradual index distribution within the fiber [69]. The extension of the latter approach is the nano-engineered fiber core [70] that uses complex refractive index distribution with the nanometric features to reach the desired functionalities. The disadvantage of the GRIN lens is that, being composed of dielectrics, it cannot overcome the diffraction limit.

## 6.2. Plasmonic lens

Using plasmonic effects in metal-dielectric zone plates is quite new since fabrication of fine metallic structures has become possible only recently. The plasmons, excited in the concentric grating by an incident light wave, contribute to the energy transfer to the central ring. The wave constructive interference condition should be satisfied not only for the diffracted light waves, but also for the plasmons. A comprehensive review of the plasmonic lenses can be found in [71]. An interesting property of the plasmonic Fresnel lens is that different wavelengths have different focal points. That gives opportunity to use such lenses for spectroscopy purposes.

The focusing effect of a plasmonic lens was experimentally shown in 2002 by Lezec et al. [72]. The theory and performance dependence on the geometry were discussed in the works [73–75]. A lens consisting of concentric rings of holes (the so-called nanopinhole lens) was numerically analyzed in [76]. Such lens allows for high transmission at $\lambda = 550$ nm with focal spot of 250 nm, that is just a bit less than a half of the wavelength. A subwavelength focusing of the $\lambda = 532$ nm monochromatic light into the spot of $\lambda/10$ with the transmittivity of 30% was numerically shown in [77]. Such high transmittivity is reached by adding a resonator to the entrance of the lens. Experimental investigation of the subwavelength focusing was also conducted [78]. Focusing of the visible light wave with $\lambda = 633$ nm to a subwavelength spot with the diameter 468 nm was reported.

As can be seen from the abovementioned results the plasmonic lens can provide the subwavelength focusing, but practically the focal spot is not much smaller comparing to a dielectric lens, while the transmittivity is lower due the presence of metal elements.

### 6.3. Negative refractive index lens

The seminal idea that a plane slab of a negative index material may focus light was mentioned by V. Veselago in 1968 [79]. However, the interest to the negative index material lenses exploded only at the beginning of the 21[st] century, when J. Pendry showed that a negative index slab not only focuses the propagating waves, but also enhances the evanescent waves, which contain the deep-subwavelength image details [80]. Such perfect lens works in the near-field regime and can provide the ideal image that repeats all small details of an original object. To function as a super-resolution lens the negative index material should be isotropic. Some special cases of anisotropic materials without spatial dispersion may also be suitable [81].

The negative refractive index occurs as a rule in materials that simultaneously possess negative dielectric permittivity and magnetic permeability. A comprehensive overview of the negative index metamaterials can be found in [82], [83].

A negative index slab lens can be used not only for focusing and imaging, but also as a coupler. The idea to use a flat negative index slab for coupling between two identical nano-waveguides was proposed by Degiron et al. [84]. By making a concave lens out of NIM, one can match two waveguides with different cross-section. However, to be utilized for the nanocoupler construction the negative refractive index material must have bulk isotropic optical properties with small losses. The requirement for the bulk behavior means that the effective properties of the homogenized metamaterial should not depend on the slab thickness. It is usual situation that the properties of a metamaterial depend on the thickness (number of monolayer) due to the coupling between monolayers [85] with rare exceptions only [86].

In the most cases proposed metamaterials consist of planar layers, since their fabrication is based on the planar technology. That results in the optical anisotropy. To overcome such drawback several isotropic NIMs designs were proposed [87–93], but the material parameters needed to obtain the desired resonances in the optical regime can hardly be found in nature, thus limiting the application of such design to the microwave region. Moreover, the unit cell of most metamaterials is not very small comparing with the wavelength of light (usually it is in the order of $\lambda/10$-$\lambda/4$) that leads to spatial dispersion and hence deterioration of the effective properties introduction. So even the highest possible cubic symmetry of the unit cells and placement of unit cells in a cubic lattice do not ensure optical isotopy of negative index metamaterials [94].

The losses in a negative index metamaterial coupler are an important issue, since they reduce the coupling efficiency and decrease the spatial resolution. This is why a lot of efforts is applied at the moment to compensate the losses with gain material [95–97] or by switching for better plasmonic materials [98], [99]. Moreover, as it is shown in recent work [100], although the subwavelength resolution is possible in metamaterial based lenses, practical focusing beyond the diffraction limit is challenging and when designing such device one should consider "granularity, degree of isotropy and transverse size of the metamaterial lens". Large losses, low coupling efficiency, anisotropy and fabrication difficulties prevent a practical realization of the negative index metamaterial lens nanocoupler.

### 6.4. Photonic crystal lens

The negative refraction phenomenon may occur not only in negative index materials but also in photonic crystals typically at frequencies close to the bandgap edge [101]. This effect can be exploited to make a photonic crystal slab working for light focusing. For example, in the theoretical work [102] there was reported that a photonic crystal in the so-called canalization regime provides the subwavelength focusing down to a spot of $\lambda/6$. To increase the transmission through a photonic crystal lens an antireflection coating can be applied [103]. Experimental realization of an *InGaAsP/InP* photonic crystal lens for $\lambda = 1.5$ μm was reported in the work [104]. They managed to obtain a focal spot of the area $0.12\lambda^2$ that overcomes the diffraction limit. Similar results for $\lambda = 1.55$ μm in a *InP/InGaAsP/InP* photonic crystal lens were obtained in the work [105], where a focal spot of $0.38\lambda$ size was demonstrated.

### 6.5. Hyperlens

Another solution for the subwavelength focusing involves a material with the hyperbolic dispersion, a so-called indefinite medium [106], [107] that simultaneously possesses positive and negative principle components of the permittivity tensor. This results in a hyperbolic isofrequency diagram, not circular or elliptical ones as in the case of conventional dielectric with positive permittivity. An important advantage of the hyperbolic material is that for a certain permittivity tensor it allows propagating waves with any values of the tangential component of the wavevector.

A medium with the hyperbolic dispersion is anisotropic by definition. To reach positive and negative permittivity for different directions one may use either metallic wires or a metal-dielectric multilayer stack. In case of wires the effective permittivity can be negative for electric field polarization along the wires, while for the polarization perpendicular to the wires the relevant principle value of the permittivity tensor is positive. In case of the metal-dielectric stack, the electric field polarized parallel to the metallic plates experiences negative dielectric response, while the permittivity associated with the perpendicular polarization is positive. A detailed theory of the multilayer and wire-medium hyperlens can be found in the works [108], [109].

The first theoretical work on a wire medium hyperlens [110] showed a subwavelength resolution $\lambda/6$ for $\lambda = 1.5$ μm. In another theoretical work [109] the $\lambda/10$ resolution was achieved in the infrared range. The same device were successfully simulated and experimentally validated in the microwave range [111–113]. In the work [113] an impressive $\lambda/15$ resolution was experimentally demonstrated.

Arranging metal wires in the tapering-up-like manner, not only the 1:1 image transfer, but also the image magnification can be achieved. An analysis of the homogeneous medium approximation eligibility for the wire medium superlens was conducted in work [114]. In the work [115] a lens for color imaging in the visible range is proposed. The idea is to engineer the wavelength selective response introducing gaps in metallic nanowires. The experimental demonstration of a wire lens for the optical range [116] and for telecom $\lambda = 1.55$ μm has been recently shown [117]. The resolution measured with a scanning near-field optical microscope was about $\lambda/4$.

A flat metal-dielectric stack can transfer the subdiffraction image as a near-field lens. Using cylindrical or spherical multilayer system the image may be magnified up to above the diffraction limit. Such magnified image may be registered with an optical microscope afterwards. So the hyperlens can serve as an addition to the standard optical microscope or photolithographic system improving the resolution below the diffraction limit. The first theoretical work [118] showed the resolution of $\lambda/4.5$. The hyperlens theory was described in paper [108]. The experimental realization of the lensing effect for the ultraviolet light $\lambda = 365$ nm, which is a standard wavelength for the optical lithography, with the resolution less than $\lambda/4$ was presented [119]. In theory the resolution was pushed further significantly. Ultraviolet light designs providing $\lambda/18$ [120] and $\lambda/60$ [121] resolution have been recently proposed.

Despite of the fantastic resolution, the hyperlens typically has low transmission. Using the Fabry-Perot effect it is possible to increase the transmittivity values. For example, in the work [122] the three-layer thick hyperlens showed 50% transmission. However, the magnification of such hyperlens was very low due to the small ratio of the outer and inner radii - about 1.25. Making the thickness of the hyperlens thicker, one would lose a lot in transmission. So, when designing a hyperlens one should find a trade-off between tight focusing and high transmission.

**Summary**

An imaging device such as lens does not necessarily provide high coupling efficiency since the primary goal of imaging is focusing of light, regardless of the transmittivity. The dielectric lens can give high transmittivity (close to 100%) but the focal spot size is diffraction limited. A photonic crystal lens can indeed provide focusing at a specific frequency. However, the resolution of such lens is not much better than the diffraction limit allows. The plasmonic lens can provide a better spatial resolution with the cost of lower transmittivity. The negative index metamaterial lens is still far from the practical realization. From the resolution point of view, the hyperlens is the best. Theoretically it can provide the resolution as small as $\lambda/60$. However, the transmittivity through the hyperlens is not high due to the employment of a metal-dielectric multilayer stack or metallic nanowires, which hinder its coupling applicability.

## 7. Scatterer coupler

The main idea behind the scatterer coupler is that there are single or multiple particles that first capture the radiation from the free space and then launch it into the waveguide. Based on the geometrical placement and material we singled out the following types :

1. Antenna coupler (Fig. 9(a)).

2. Grating coupler (Fig. 9(b)).

3. Random scatterers (Fig. 9(c)).

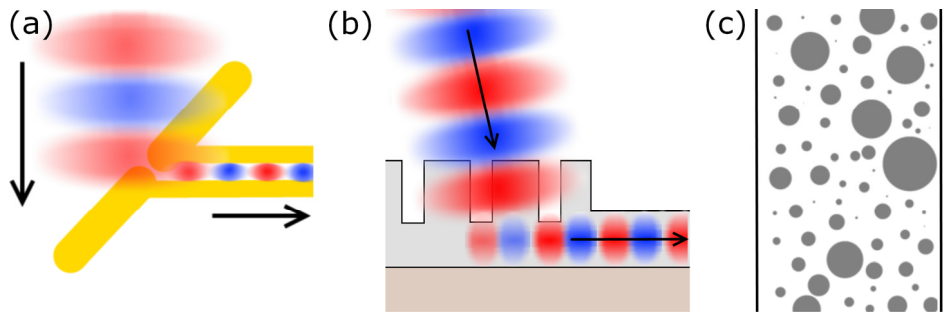

Figure 9. Types of scatterer couplers: (a) antenna, (b) grating, (c) random scatterers.

## 7.1. Antenna coupler

Accordingly to the definition [123], [124], an antenna is *a device that converts a free propagating radiation into the localized power and vice versa*. For example, television antennas capture the waves propagating in air and transform them into an electrical current. In other words, an antenna is a coupler that matches together the impedances of free space and a waveguide. This application of antennas account for more than one hundred years, since Hertz, Popov, and Markoni invented the principles of the radio transmission. Now radio and microwave antennas are well-known engineering systems.

In principle, according to the definition, some nanocouplers of other types, for example, a lens coupler, can also be called antennas. In this section, however, we limit ourselves to the traditional geometrical antenna configurations, which consist of a single of multiple metallic particles, specially tuned for accepting electromagnetic radiation.

Optical nanoantennas drew the attention of many research groups [125–135]. Nanoantennas can be considered as the optical analogue of the microwave and radio antennas [125]. An incident electromagnetic wave excites charges oscillations along the metallic antenna, which are in the optical range nothing else than localized surface plasmons, and then couples to a mode of the connected waveguide.

Despite a lot of similarities, there are essential differences between radio and plasmonic antennas [136]:

1. Metals are not such good conductors in the optical range as on the radio frequencies. Their permittivity is dispersive. It can be approximated by the Drude or more accurate Drude-Lorentz formulas [3] accounting for interband transitions in noble metals.

2. A typical penetrations depth is several dozens of nanometers [3] and that is very important for metallic nanostructures of comparable size.

3. The usual condition for the resonance of radiofrequency antennas is that the length of the antenna must be equal to the integer of a half of the wavelength. This condition is not satisfied for the optical antennas and should be corrected [128], [137], [138].

Plasmonic nanoantennas have attracted huge attention in the recent years because of their ability to concentrate light in the tiny gaps and significantly enhance light intensity [126], [127], [133], [138]. A comprehensive overview of the nanoantennas theory and applications can be found in [123], [136], [139].

Despite of the fact that from the very beginning antennas served for coupling to a waveguide (transmission line), the application of an antenna for optical nanocoupling was proposed only recently [140–146]. The reason for that are mainly the difficulties of ultrafine metallic structures fabrication (feature size on the order of 10-100 nm).

The advantage of the nanoantenna is that it is very compact. Moreover, the directivity of the antenna coupler can be tuned by design, thus enabling the maximal coupling efficiency at any desired angle of incidence. The first nanoantenna couplers analysis showed theoretically the coupling efficiency of 10% [140] for $\lambda = 1.55$ µm and 28% [141] for $\lambda = 830$ nm for a focused Gaussian spot. Experimentally measured coupling efficiency of 15% for $\lambda = 1.55$ µm was demonstrated in the work [144]. It was shown that the coupling efficiency can be increased at least by two times by applying additional reflectors and arranging nanoantennas in a parallel or serial array [145].

## 7.2. Grating coupler

An array of nanoantennas is closely related to the diffraction grating employment as a coupler. However, the difference is that each nanoantenna has a specific length designed to be in resonance with an incident electromagnetic wave, while in the diffraction grating each line can be very long. The lines of the diffraction grating lie on top of the waveguide and scatter light into the waveguide. It is of outmost importance that the scattered radiation from each line contributes constructively to the wave propagating in the waveguide. In other words, the role of the grating is to match the tangential wavevector component $k_t$ of the incident wave with the waveguide propagation constant $\beta$. The grating with a specific period $P$ allows light to diffract to a diffraction order $m$ such that $\beta = k_t + 2\pi m/P$.

A diffraction grating is typically employed for coupling to a silicon waveguide in the vertical coupling configuration. The spot size from a single mode fiber is usually about 10 µm. The wave should not be strictly perpendicular to the waveguide, but incident at a small angle (about 5 degrees) to provide the wavevector matching. The maximal coupling efficiency of 60-70% was reported [147], [148]. There was also claimed a high efficiency (96%) vertical coupler based on the subwavelength grating [149]. Making the grating lines concentric and adding tapering it is possible to couple and focus light simultaneously. For example, a 40 µm-size focusing grating coupler with 25% coupling efficiency at $\lambda$=2.75 µm was demonstrated in the work [150].

A grating coupler can also be used for the surface plasmon polaritons waveguide excitation. The coupling efficiency up to 68% was theoretically shown [51], [151–153]. The grating can be designed to be polarization independent [154]. It can also combine the functions of a coupler and a nonlinear higher order harmonics generator [155].

## 7.3. Random scatterers

As we mentioned before the nanocoupler can be understood as a mode transforming device. In principle we can designate two extreme cases for mode conversion: evolution, i.e. a careful adiabatic compression towards mode profile matching – this is realized by long tapered fibers; and revolution, i.e. complete mode structure destruction and then construction another mode in the same way as a new building can be built from the bricks of a ruined house. This analogy would mean introduction of a set of random scatterers. The photons coming from the first waveguide experience multiple scattering and statistically some of them can couple to a mode of the second waveguide.

It is clear that the coupling efficiency of such random material coupler cannot be high due to the random nature of the photon scattering process. However, there was recently shown that under certain circumstances a disordered medium can work for light focusing [156].

**Summary**

The antenna nanocoupler is a natural transition of a standard microwave approach for coupling an electromagnetic wave to an optical (plasmonic) waveguide. There is a theoretical 50% limit of the coupling efficiency of the antenna systems [157] due to re-radiation of the captured power back into free space. Practically, the nanoantennas can exhibit the coupling efficiency close to the theoretical limit, being constrained only by optical losses and fabrication imperfections. Nevertheless, the main advantage of the nanoantenna coupler is that it is the most compact among all the nanocoupling solutions. Diffraction gratings are very efficient vertical coupling solutions providing the coupling efficiency up to 70%. Their disadvantage is relatively large size (more than 10 µm). The random scatterers have very low coupling efficiency due to stochastic scattering of light into the desired waveguide mode, so they are hardly suitable for the nanocoupling applications.

## 8. Other solutions

In this section we included all other ideas that do not fall into the abovementioned categories. Such coupling ideas are:

1. Transformation optics coupler.

2. Topology optimization designed coupler.

### 8.1. Transformation optics coupler

Inspired by the metamaterials possibilities of obtaining whatever permittivity and permeability values the field of transformation optics has recently emerged [158–160]. The transformation optics solves the problem of determination of the spatial permittivity and permeability distribution that provides a required wave propagation trajectory. For example, in case of invisibility cloaking engineering it is required that light rays pass around an object not interacting with it [161].

A problem of coupling can be expressed in the language of the transformation optics. To design a nanocoupler is to determine such spatial distribution of permittivity and permeability that provides impedance matching and squeezing of the light from a thick waveguide to a thin one. In some sense the GRIN lens is also an example of the transformation optics application. Some designs for squeezing light [162] or light concentration [163], [164] were proposed.

Very often, however, the transformation optics designs require unusual values of permittivity and permeability that are not realistic even with the help of metamaterials (for example, diverging permittivity and permeability values without losses or extremely large anisotropy).

### 8.2. Topology optimization

Another approach to the coupler design is to select from the very beginning the realistic material properties (in the simplest case, of two materials) and to determine the spatial distribution (topology) of the materials that gives the largest coupling efficiency. With the efficient numerical algorithms one has no need to go deep into physical consideration when designing the nanocoupler. Such approach is called *topology optimization* [165]. The optimized structures usually have very weird shapes [166]. Setting some constraints on the geometrical size of the fine features one can design an efficient coupler that is reasonable for fabrication [167].

**Summary**

The transformation optics devices very often require unrealistic material properties. In contrary to the transformation optics the topology optimization starts from the realistic material properties and then finds the necessary geometry. We should admit that both of these approaches can be applied to almost any coupler in the sections 4-7. Therefore we should better say that these are not independent physical approaches, but rather useful design methodologies.

## Conclusions

In this paper we have analyzed various physical principles that can be used for coupling light from an optical fiber or free space to a nanosized waveguides. The range of approaches is very broad, so we divided the subject into four classes (tapered waveguide coupler, direct coupler, lens coupler and scatterer coupler). The most important features of each approach are summarized in Table 1.

Table 1. Comparison of various nanocoupling approaches.

| Approach | Coupling efficiency: low (<10%), medium (10-50%), high (>50%) | Size: compact (<10 μm) or large (>10 μm) | Subwavelength coupling or focusing | Lateral or vertical coupling | References |
|---|---|---|---|---|---|
| 4.1. Tapered dielectric | high | both | no | Lateral | [18–25] |
| 4.2. Tapered metal | medium and large | compact | yes | Lateral | [26–39] |
| 4.3. Tapered metamaterial | low and medium | compact | yes | Lateral | [40–42] |
| 5.1. Next to another | high | both | yes | Lateral | [43–53] |
| 5.2. End-fire | high | compact | no | Lateral | [54–60] |
| 5.3. Resonant stub | high | compact | yes | Lateral | [11], [61–64] |
| 6.1. Dielectric lens | high | large | no | Lateral | [66], [67], [69], [70] |
| 6.2. Plasmonic lens | medium | compact | yes | Lateral | [71–78] |
| 6.3. Negative index lens | N/A | N/A | yes | Lateral | [79], [80], [84], [100] |
| 6.4. Photonic crystal lens | N/A | compact | yes | Lateral | [101–105] |
| 6.5. Hyperlens | low or medium | compact | yes | Lateral | [107–122] |
| 7.1. Antenna | medium | compact | yes | Both | [140–146] |
| 7.2. Grating | high | large | no | Vertical | [147–155] |

| 7.3. Random scatterers | low | N/A | yes | Both | [156] |

The most compact solution for the nanocoupling is the antenna coupler. The most efficient are the tapered waveguide and the grating coupler combined with the directional coupler. The hyperlens gives a good trade-off between subdiffraction imaging and transmission. The designs that use negative index materials are lossy and therefore can hardly be used for the nanocoupler at the moment. A discovery of new plasmonic materials with smaller losses or optical losses compensation with gain can probably make the latter approaches useful for light coupling.

We see the future of the nanocouplers mostly in their technical improvement. This includes a search for better materials, optimization of the designs and fabrication technologies. The abovementioned coupling approaches can also be the building blocks of more advanced photonic devices. For example, tapering the directional slot waveguides coupler and filling the slots with a nonlinear material [168] gives a new nanophotonic functionality and can be used for all-optical switching. Another direction is the transfer of the optical coupling approaches to other fields of physics, for example, to acoustics or surface waves.

## Acknowledgements

The authors acknowledge the members of the Metamaterials group at DTU Fotonik for useful discussions and the financial support from the Danish Council for Technical and Production Sciences through the projects GraTer (11-116991), NIMbus and THzCOW.